\documentclass[reprent,aps,prb,twocolumn,
superscriptaddress,nofootinbib]{revtex4-2}
\usepackage{amsmath,amssymb}
\usepackage{graphicx}% Include figure files
\usepackage{bm}% bold math
\usepackage{multirow}
\usepackage{hyperref}
\usepackage{braket}

\usepackage{color}

\allowdisplaybreaks[2]

\begin{document}

\title{
Field-induced transformation between triangular and square skyrmion crystals\\ in a tetragonal polar magnet
}
\author{Satoru Hayami}
\email{hayami@phys.sci.hokudai.ac.jp}
\affiliation{Graduate School of Science, Hokkaido University, Sapporo 060-0810, Japan}

\begin{abstract}
Magnetic skyrmions with topologically nontrivial spin textures form a variety of periodic structures depending on microscopic interactions and lattice symmetry. 
We theoretically investigate a transformation between triangular and square skyrmion crystals against an external magnetic field in a polar tetragonal magnet. 
By performing the simulated annealing for a classical spin model, we show that the competition of the Dzyaloshinskii-Moriya interaction at multiple wave vectors is a key ingredient in inducing the structural transition in terms of the skyrmion crystals. 
The present results indicate the importance of magnetic frustration in momentum space as the origin of exotic topological phase transitions. 
\end{abstract}

\maketitle

\section{Introduction}

Noncoplanar spin textures have drawn extensive interest as a new quantum magnet in condensed matter physics, since they bring about fascinating physical phenomena owing to emergent electromagnetic fields arising from the quantum mechanical Berry phase. 
Various unconventional transport phenomena induced by noncoplanar spin textures have been explored, such as the topological Hall effect~\cite{Loss_PhysRevB.45.13544, Ye_PhysRevLett.83.3737, Ohgushi_PhysRevB.62.R6065, Shindou_PhysRevLett.87.116801, tatara2002chirality, Haldane_PhysRevLett.93.206602, Nagaosa_RevModPhys.82.1539, Zhang_PhysRevB.101.024420, takagi2023spontaneous}, nonreciprocal transport~\cite{Hayami_PhysRevB.101.220403, hayami2021phase, Hayami_PhysRevB.106.014420, Hayami_doi:10.7566/JPSJ.91.094704, Eto_PhysRevLett.129.017201}, and magneto-optical effect~\cite{feng2020topological}. 
Especially, a magnetic skyrmion crystal (SkX) is a typical example of exhibiting the above physical phenomena, which has been found in magnetic materials under distinct lattice structures since its discovery in 2009~\cite{Muhlbauer_2009skyrmion, nagaosa2013topological, Tokura_doi:10.1021/acs.chemrev.0c00297}. 

The SkX is formed by a periodic array of magnetic skyrmion as particle-like topological spin textures. 
Among them, most of the SkXs in materials are characterized by the hexagonal packed structure~\cite{Binz_PhysRevB.74.214408, Yi_PhysRevB.80.054416, Muhlbauer_2009skyrmion,yu2010real, yu2011near}, although some of the materials host the tetragonal packed one~\cite{khanh2020nanometric, Yasui2020imaging, khanh2022zoology}.  
These SkX phases often emerge by introducing an external magnetic field when the single-$Q$ helical spiral state is stabilized at zero field; the spin texture in the SkX is expressed as a superposition of multiple spiral waves at different ordering wave vectors. 
Then, the SkX is replaced by different magnetic phases, such as a single-$Q$ conical spiral state and a fully polarized state, with a further increase of the magnetic field. 
In general, the SkX phase appears as only one phase against the magnetic field. 

Multiple SkX phases with different packed structures have recently been observed in several materials by changing the magnetic field and temperature. 
One of the examples is the archetypal skyrmion-hosting chiral compound MnSi~\cite{Muhlbauer_2009skyrmion}, where a transition from a square SkX (S-SkX) to a triangular SkX has been found as a metastable state after thermal quenching~\cite{nakajima2017skyrmion}; similar structural transitions of the SkX have also been found in other noncentrosymmetric magnets, such as Co$_8$Zn$_8$Mn$_4$~\cite{karube2016robust, Karube_PhysRevB.102.064408, Henderson_PhysRevB.106.094435}. 
Another example is a centrosymmetric tetragonal magnet EuAl$_4$. 
In contrast to MnSi and Co$_8$Zn$_8$Mn$_4$, two types of SkXs, which were identified as the S-SkX and a rhombic SkX, have been observed as thermodynamics phases by changing the magnetic field~\cite{takagi2022square, Zhu_PhysRevB.105.014423, hayami2023orthorhombic, Gen_PhysRevB.107.L020410}. 
In this compound, the model calculations have shown that the competition of the interactions in momentum space that arise from the Ruderman-Kittel-Kasuya-Yosida (RKKY) interaction mediated by itinerant electrons~\cite{Ruderman, Kasuya, Yosida1957} or frustrated exchange interactions plays an important role~\cite{takagi2022square, hayami2023orthorhombic, Gen_PhysRevB.107.L020410}. 
More recently, such structural transition of the equilibrium SkX phases has also been observed in a polar magnet EuNiGe$_3$, where the long-range Dzyaloshinskii-Moriya (DM) interaction~\cite{dzyaloshinsky1958thermodynamic,moriya1960anisotropic}, as well as the RKKY interaction, contributes to the emergence of the multiple SkXs~\cite{singh2023transition}. 
These experimental findings indicate that exotic topological phase transitions including multiple SkX phases are brought about by magnetic frustration in momentum space irrespective of the spatial inversion symmetry~\cite{hayami2021topological}.  

In the present study, we numerically investigate the instability toward multiple SkXs against the external magnetic field, which is induced by the competition between the long-range RKKY and DM interactions. 
For that purpose, we analyze an effective spin model on a polar square lattice, which is derived by extracting the important interactions in momentum space. 
We construct the magnetic phase diagram at low temperatures by performing the simulated annealing. 
As a result, we find that the competing interactions in momentum space give rise to phase transitions among the different types of SkX phases. 
We show that the structural transition from the distorted triangular SkX (DT-SkX) to the S-SkX occurs as found in EuNiGe$_3$~\cite{singh2023transition}, and that from the distorted S-SkX (DS-SkX) to the S-SkX occurs as found in EuAl$_4$~\cite{takagi2022square, hayami2023orthorhombic} depending on the degree of frustration in momentum space. 
Our results provide an important essence to induce multiple SkX phases, which will be useful for further exploration of topological spin textures in noncentrosymmetric magnets. 

The organization of this paper is as follows. 
In Sec.~\ref{sec: Model and method}, we introduce a spin model on a polar square lattice. 
We discuss the role of the momentum-resolved interaction. 
We also present a numerical method based on the simulated annealing. 
In Sec.~\ref{sec: Results}, we present the magnetic phase diagram, and then, we discuss the phase transitions between the SkXs in detail. 
Section~\ref{sec: Summary} is devoted to a summary of the present paper.

\section{Model and method}
\label{sec: Model and method}

\begin{figure}[tb!]
\begin{center}
\includegraphics[width=1.0\hsize]{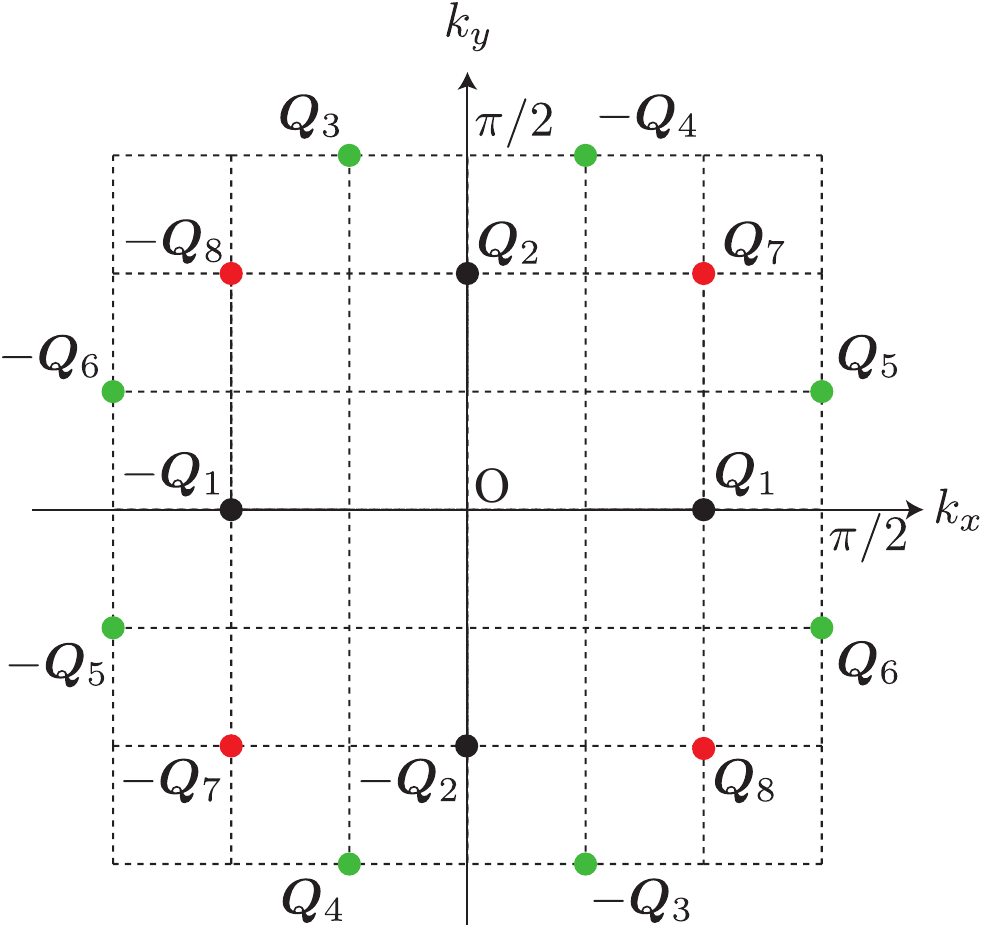} 
\caption{
\label{fig: Qvec} 
Ordering wave vectors that give the dominant contributions in the momentum-resolved interactions.
The coupling constants in the $(\pm \bm{Q}_1, \pm \bm{Q}_2)$, $(\pm \bm{Q}_3, \pm \bm{Q}_4, \pm \bm{Q}_5, \pm \bm{Q}_6)$, and $(\pm \bm{Q}_7, \pm \bm{Q}_8)$ components of the momentum-resolved interactions are given by $J_1$, $J_2$, and $J_3$ in the model in Eq.~(\ref{eq: Ham}), respectively. 
}
\end{center}
\end{figure}

We consider a spin model on a two-dimensional square lattice with the lattice constant $a=1$ under the polar point group $C_{4v}$, whose Hamiltonian is given by 
\begin{align}
\label{eq: Ham}
\mathcal{H}=&-\sum_{\nu}  \left[J_{\bm{Q}_\nu} \bm{S}_{\bm{Q}_\nu} \cdot  \bm{S}_{-\bm{Q}_\nu} 
+ i \bm{D}_{\bm{Q}_\nu} \cdot (\bm{S}_{\bm{Q}_\nu} \times  \bm{S}_{-\bm{Q}_\nu})\right] \nonumber \\
&-H\sum_i  S^z_i, 
\end{align}
where the first term stands for the $\bm{Q}_\nu$ component of momentum-resolved interaction; $\bm{S}_{\bm{Q}_\nu}=(S^x_{\bm{Q}_\nu}, S^y_{\bm{Q}_\nu}, S^z_{\bm{Q}_\nu})$ is the Fourier transform of the localized classical spins $\bm{S}_i=(S_i^x, S_i^y, S_i^z)$ with the length $|\bm{S}_i|=1$. 
The former term represents the Heisenberg-type exchange interaction with the coupling constant $J_{\bm{Q}_\nu}(=J_{-\bm{Q}_\nu})$, while the latter term represents the DM interaction with the coupling constant $\bm{D}_{\bm{Q}_\nu} (= -\bm{D}_{-\bm{Q}_\nu})$. 
The second term stands for the Zeeman coupling to take into account the effect of an external magnetic field along the $z$ direction. 

In order to investigate the competition of the interactions in momentum space, we consider the momentum-resolved interactions at specific ordering wave vectors. 
We consider sixteen interaction channels: $\pm \bm{Q}_1=\pm (\pi/3,0)$, $\pm \bm{Q}_2=\pm (0, \pi/3)$, $\pm \bm{Q}_3=\pm (-\pi/6, \pi/2)$, $\pm \bm{Q}_4=\pm (-\pi/6, -\pi/2)$, $\pm \bm{Q}_5=\pm (\pi/2, \pi/6)$, $\pm \bm{Q}_6=\pm (\pi/2, -\pi/6)$, $\pm \bm{Q}_7=\pm (\pi/3, \pi/3)$, and $\pm \bm{Q}_8=\pm (\pi/3, -\pi/3)$, which are schematically shown in Fig.~\ref{fig: Qvec}. 
$\bm{Q}_1$ and $\bm{Q}_2$ ($\bm{Q}_7$ and $\bm{Q}_8$) are symmetry-related wave vectors, which are connected by both fourfold rotational and mirror operations. 
Similarly, $\bm{Q}_3$--$\bm{Q}_6$ are symmetry-equivalent to each other, which are connected by the fourfold rotational operation or mirror operation; for example, $\bm{Q}_3$ is connected to $-\bm{Q}_5$ by the fourfold rotational operation, while it is connected to $\bm{Q}_4$ by the vertical mirror operation including the $k_x$ axis. 
Accordingly, $J_{\bm{Q}_1}=J_{\bm{Q}_2}=J_1$, $J_{\bm{Q}_3}=J_{\bm{Q}_4}=J_{\bm{Q}_5}=J_{\bm{Q}_6}=J_2$, and $J_{\bm{Q}_7}=J_{\bm{Q}_8}=J_3$ for the isotropic exchange interaction and $|\bm{D}_{\bm{Q}_1}|=|\bm{D}_{\bm{Q}_2}|=D_1$, $|\bm{D}_{\bm{Q}_3}|=|\bm{D}_{\bm{Q}_4}|=|\bm{D}_{\bm{Q}_5}|=|\bm{D}_{\bm{Q}_6}|=D_2$, and $|\bm{D}_{\bm{Q}_7}|=|\bm{D}_{\bm{Q}_8}|=D_3$ for the DM interaction. 
The DM vector in polar crystals is perpendicular to both ordering wave vectors and out-of-plane $k_z$ axis for high-symmetric wave vectors; $\bm{D}_{\bm{Q}_\nu}  \parallel  \hat{\bm{k}}_z \times\bm{Q}_\nu$ for $\nu=1,2,7,8$, where $\hat{\bm{k}}_z$ represents the unit vector along the $k_z$ direction. 
Although the in-plane direction of the DM vector at low-symmetric wave vectors for $\nu=3,4,5,6$ is not necessarily to take the perpendicular direction to $\hat{\bm{k}}_z \times\bm{Q}_\nu$ from the symmetry viewpoint~\cite{Yambe_PhysRevB.106.174437}, we also set $\bm{D}_{\bm{Q}_\nu} \parallel \hat{\bm{k}}_z  \times\bm{Q}_\nu$ for simplicity.  
We neglect the symmetric anisotropic exchange interaction and dipolar interaction, which can be a source of the SkXs~\cite{amoroso2020spontaneous, Hayami_PhysRevB.103.024439, Hayami_PhysRevB.103.054422, amoroso2021tuning, Hirschberger_10.1088/1367-2630/abdef9, Utesov_PhysRevB.103.064414, Wang_PhysRevB.103.104408, Kato_PhysRevB.104.224405,  Nickel_PhysRevB.108.L180411}. 

One of the microscopic origins of the momentum-resolved interactions, $J_{\bm{Q}_\nu}$ and $\bm{D}_{\bm{Q}_\nu}$, is the long-range RKKY interaction in itinerant electron systems~\cite{Hayami_PhysRevLett.121.137202}. 
In this case, the important $\bm{Q}_\nu$ channel in the interaction is determined by the nesting property of the Fermi surface. 
Another microscopic origin of $J_{\bm{Q}_\nu}$ and $\bm{D}_{\bm{Q}_\nu}$ is the frustrated exchange interaction in real space~\cite{Wang_PhysRevB.103.104408, Hayami_PhysRevB.105.174437}. 

Let us discuss the geometrical relation among $\bm{Q}_1$--$\bm{Q}_8$ in Fig.~\ref{fig: Qvec}. 
The wave vectors $\bm{Q}_7$ and $\bm{Q}_8$ correspond to high-harmonic wave vectors of $\bm{Q}_1$ and $\bm{Q}_2$: $\bm{Q}_7=\bm{Q}_1 + \bm{Q}_2$ and $\bm{Q}_8=\bm{Q}_1-\bm{Q}_2$. 
Such a relation can lead to the instability toward the S-SkX when the interaction at $\bm{Q}_7$ and $\bm{Q}_8$, i.e., $J_3$ is comparable to that at $\bm{Q}_1$ and $\bm{Q}_2$, $J_1$~\cite{Hayami_PhysRevB.105.174437, hayami2022multiple, hayami2023widely}. 
On the other hand, by combining $(\bm{Q}_3, \bm{Q}_4, \bm{Q}_5, \bm{Q}_6)$ and $(\bm{Q}_7, \bm{Q}_8)$, the wave vectors at $\bm{Q}_3$--$\bm{Q}_8$ can satisfy the relation as $\bm{Q}_4+\bm{Q}_5-\bm{Q}_8=\bm{0}$ and $\bm{Q}_3+\bm{Q}_6-\bm{Q}_7=\bm{0}$, which often leads to the instability toward the DT-SkX~\cite{hayami2021field, Hayami_doi:10.7566/JPSJ.91.093701}. 
Thus, the competing interactions among the wave vectors $\bm{Q}_1$, $\bm{Q}_2$, $\bm{Q}_7$, and $\bm{Q}_8$ tend to favor the S-SkX, while those among $\bm{Q}_3$, $\bm{Q}_4$, $\bm{Q}_5$, $\bm{Q}_6$, $\bm{Q}_7$, and $\bm{Q}_8$ tend to favor the DT-SkX. 

With this tendency in mind, we set the model parameters $(J_1, J_2, J_3, D_1, D_2, D_3, H)$ in Eq.~(\ref{eq: Ham}) as follows. 
We set $J_3=1$ as the energy unit of the model, and consider it the dominant interaction in the model. 
Since we examine the magnetic instability under the competing interactions in momentum space, we set $J_2=0.95$, and deal with $J_1$ as a variable parameter ($0 \leq J_1 \leq 1$). 
Thus, the interaction at $(\bm{Q}_3, \bm{Q}_4, \bm{Q}_5, \bm{Q}_6)$ gives the second largest contribution for $J_1 < 0.95$, while that at $(\bm{Q}_1, \bm{Q}_2)$ gives the second largest contribution for $J_1 > 0.95$. 
For the DM interaction, we parametrize $D_1=(J_1/J_3)D$, $D_2=(J_2/J_3)D$, and $D_3=D$ with $D=0.2$. 
We change $H$ as a variable parameter. 

While changing $J_1$ and $H$, we construct a magnetic phase diagram at low temperatures. 
We adopt a numerical method based on the simulated annealing following the manner in Ref.~\cite{hayami2020multiple}. 
We start a simulation from high temperatures $T_0=$1--5, where the initial spin configuration is taken at random. 
Then, we gradually reduce the temperature as $T_{n+1}=0.999999 T_n$ to the final lowest temperature $T=0.01$, where $T_n$ is the $n$th-step temperature. 
In each temperature, we perform the local spin updates based on the standard Metropolis algorithm. 
At the final temperature $T$, we further perform $10^5$--$10^6$ Monte Carlo sweeps for measurements. 
The simulations are independently done from a set of $J_1$ and $H$. 
In the vicinity of the phase boundaries, we start the simulations from the spin configurations obtained at low temperatures. 
The following results are calculated for the system size with $N=12^2$ under the periodic boundary condition, although we confirm that the same results are obtained for $N=48^2$. 

We show several physical quantities to identify magnetic phases obtained by the simulated annealing. 
The uniform magnetization is given by 
\begin{align}
\label{eq: magnetization}
M^\eta=\frac{1}{N} \sum_i S_i^\eta, 
\end{align}
for $\eta=x,y,z$. 
The spin structure factor is given by 
\begin{align}
S^{\eta\eta}_s(\bm{q})&=\frac{1}{N}\sum_{ij}S_i^\eta S_j^\eta e^{i \bm{q} \cdot (\bm{r}_i -\bm{r}_j)}, 
\end{align}
where $\bm{r}_i$ represents the position vector at site $i$ and $\bm{q}$ represents the wave vector in the first Brillouin zone. 
We also compute the in-plane component of the spin structure factor given by $S^{\perp}_s(\bm{q})=S^{xx}_s(\bm{q})+S^{yy}_s(\bm{q})$. 
Finally, the scalar spin chirality is given by 
\begin{align}
\chi^{\rm sc}&= \frac{1}{2 N} 
\sum_{i}
\sum_{\delta,\delta'= \pm1}
\delta \delta'
 \bar{\bm{S}}_{i} \cdot (\bar{\bm{S}}_{i+\delta\hat{x}} \times \bar{\bm{S}}_{i+\delta'\hat{y}}), 
\end{align}
where $\hat{x}$ ($\hat{y}$) represents a shift by lattice constant in the $x$ ($y$) direction. 
The SkX phase exhibits nonzero $\chi^{\rm sc}$.

\section{Results}
\label{sec: Results}

\begin{figure}[t!]
\begin{center}
\includegraphics[width=1.0\hsize]{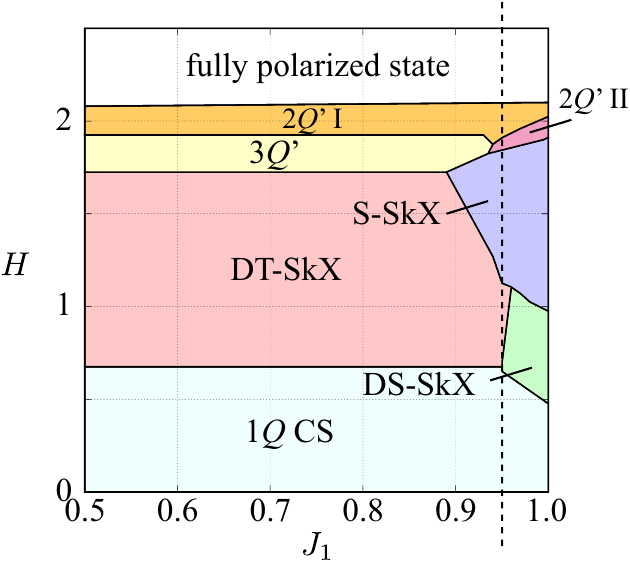} 
\caption{
\label{fig: PD} 
Magnetic phase diagram of the model in Eq.~(\ref{eq: Ham}) on the square lattice with the DM interaction, which is obtained by the simulated annealing. 
1$Q$, 2$Q$, and 3$Q$ denote the single-$Q$, double-$Q$, and triple-$Q$ states, respectively, where $Q'$ represents the anisotropic intensities of the spin moments at the multiple-$Q$ wave vectors. 
The vertical dashed line represents the boundary for $J_1=J_2$; for $J_1<0.95$ ($J_1>0.95$), the interaction at $\bm{Q}_3$, $\bm{Q}_4$, $\bm{Q}_5$, and $\bm{Q}_6$ is larger (smaller) than that at $\bm{Q}_1$ and $\bm{Q}_2$. 
}
\end{center}
\end{figure}

\begin{figure}[tb!]
\begin{center}
\includegraphics[width=1.0\hsize]{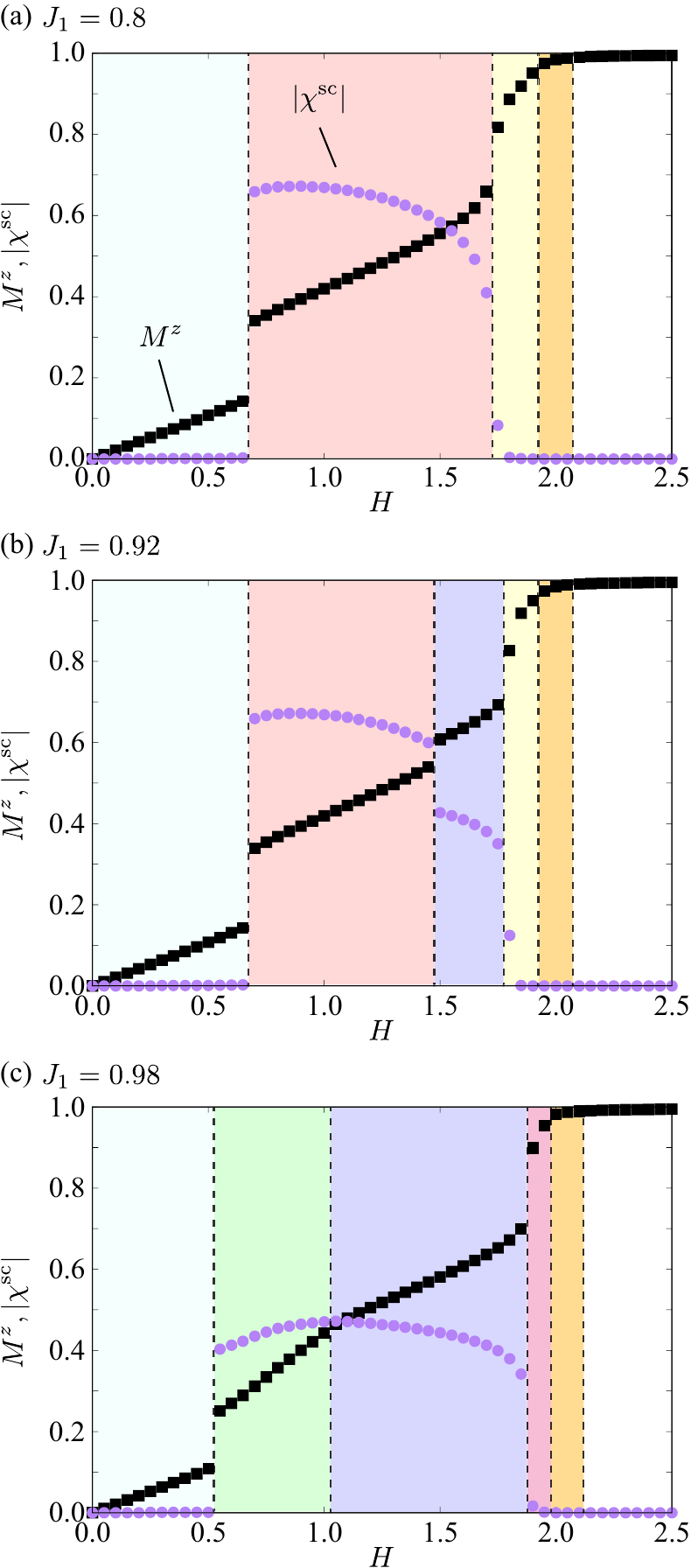} 
\caption{
\label{fig: M-H} 
$H$ dependence of the $z$ component of the magnetization $M^z$ and the scalar chirality $|\chi^{\rm sc}|$ for (a) $J_1=0.8$, (b) $J_1=0.92$, and (c) $J_1=0.98$. 
The vertical dashed lines represent the phase boundaries between different magnetic phases. 
}
\end{center}
\end{figure}

\begin{figure*}[tb!]
\begin{center}
\includegraphics[width=0.9\hsize]{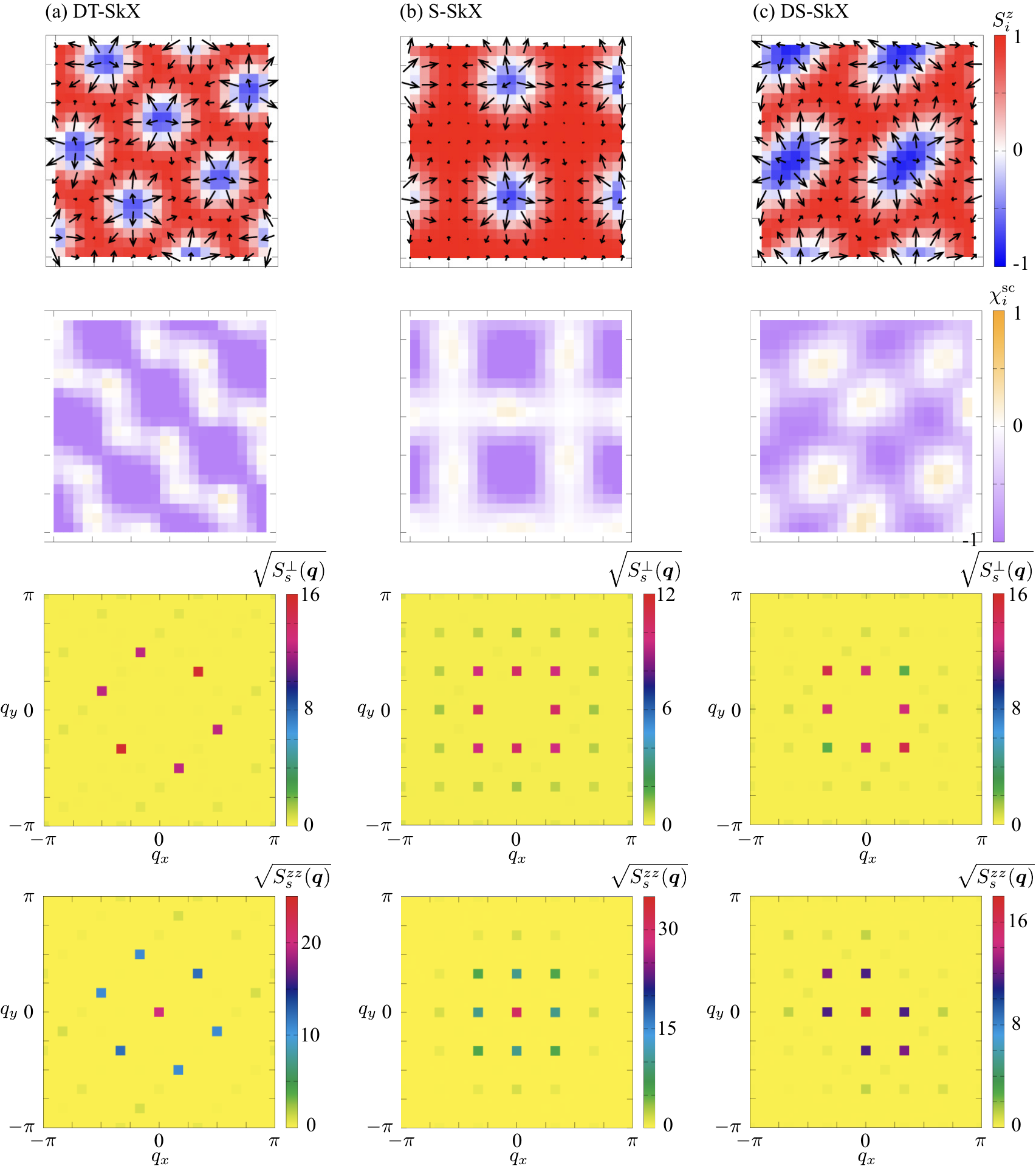} 
\caption{
\label{fig: SkX} 
(First row) Real-space spin configurations in 
(a) the DT-SkX at $J_1=0.8$ and $H=1$, 
(b) the S-SkX at $J_1=0.92$ and $H=1.6$, and
(c) the DS-SkX at $J_1=0.98$ and $H=0.8$. 
The arrows represent the direction of the in-plane spin moments and the color shows its $z$ component. 
(Second row) 
Real-space scalar spin chirality. 
(Third and fourth rows)
The square root of the (third row) $xy$ and (fourth row) $z$ components of the spin structure factor in the first Brillouin zone, which are calculated for the system size with $N=48^2$
}
\end{center}
\end{figure*}

Figure~\ref{fig: PD} shows the magnetic phase diagram obtained by the simulated annealing. 
By changing $J_1$ and $H$, we obtain eight magnetic phases with different spin and scalar chirality textures. 
Among them, we find that three SkX phases denoted as DT-SkX, S-SkX, and DS-SkX are stabilized in the phase diagram. 
We mainly discuss the details of these SkX phases. 

For $J_1 \lesssim 0.9$, five phases are stabilized, where the phase boundaries do not show the $J_1$ dependence: single-$Q$ cycloidal spiral (1$Q$ CS) state, DT-SkX, triple-$Q$ (3$Q'$) state, double-$Q$ I (2$Q'$ I) state, and fully-polarized state as $H$ increases. 
It is noted that the phase boundaries between them are unchanged up to $J_1=0$, which means that the magnetic phases for $J_1 \lesssim 0.9$ are stabilized by the competition of the interactions between $(\bm{Q}_3, \bm{Q}_4, \bm{Q}_5, \bm{Q}_6)$ and $(\bm{Q}_7, \bm{Q}_8)$ channels. 

We show the $H$ dependence of the magnetization along the field direction $M^z$ and the scalar chirality $|\chi^{\rm sc}|$ at $J_1=0.8$ in Fig.~\ref{fig: M-H}(a). 
For $H=0$, the 1$Q$ CS state appears, which is characterized by the spin density waves with $\bm{Q}_7$ or $\bm{Q}_8$ owing to $J_3 > J_1, J_2$. 
The real-space spin configuration corresponds to the cycloidal spiral, whose spiral plane is perpendicular to $\hat{\bm{k}}_z \times \bm{Q}_{7,8}$ so as to gain the energy by the DM interaction in the polar system.  
When the magnetic field increases, $M^z$ continuously increases, as shown in Fig.~\ref{fig: M-H}(a). 
Owing to the coplanar spin texture, there is no scalar spin chirality degree of freedom in this state. 

The 1$Q$ CS state turns into the DT-SkX at $H \simeq 0.68$ with jumps of $M^z$ and $\chi^{\rm sc}$, as shown in Fig.~\ref{fig: M-H}(a). 
As shown by the real-space spin configuration in the first row of Fig.~\ref{fig: SkX}(a), the skyrmion core with $S^z_i=-1$, which is located at the interstitial site~\cite{Hayami_PhysRevResearch.3.043158}, aligns in a distorted triangular-lattice way. 
Reflecting the noncoplanar spin texture, this state accompanies the uniform scalar chirality, as shown in Fig.~\ref{fig: M-H}(a) and the second row of Fig.~\ref{fig: SkX}(a). 
In momentum-space picture, this spin configuration consists of the triple-$Q$ peak structure at $\bm{Q}_3$, $\bm{Q}_6$, and $\bm{Q}_7$, where the intensity at $\bm{Q}_7$ is larger than those at $\bm{Q}_3$ and $\bm{Q}_6$, as shown in the third and fourth rows in Fig.~\ref{fig: SkX}(a); this is attributed to the interaction parameters satisfying $J_3>J_2$.  
The triple-$Q$ wave vectors are chosen by satisfying $\bm{Q}_3+\bm{Q}_6-\bm{Q}_7=\bm{0}$, which avoids the energy loss arising from the higher harmonics like $\bm{Q}_3+ \bm{Q}_6$. 
It is noted that the DT-SkX with the triple-$Q$ ordering wave vectors $\bm{Q}_4$, $\bm{Q}_5$, and $\bm{Q}_8$ also appears as an energetically degenerate state in the simulations depending on the initial spin configuration. 

By further increasing the magnetic field, the DT-SkX changes into the 3$Q'$ state at $H \simeq 1.73$.
The spin configuration in this state is similar to that in the DT-SkX, although the spins located around the skyrmion core have the polarization along the $+z$ direction. 
This indicates the cancellation tendency of the scalar chirality, which results in almost no uniform component, as shown in Fig.~\ref{fig: M-H}(a). 
The 3$Q'$ state changes into the 2$Q'$ I state at $H \simeq 1.93$, whose spin configuration is characterized by the double-$Q$ spin density waves at $\bm{Q}_7$ and $\bm{Q}_8$. 
The 2$Q'$ I state finally turns into the fully polarized state at $H \simeq 2.08$. 

When $J_1$ is comparable to $J_2$, the instability toward other SkX phases occurs, as shown in Fig.~\ref{fig: PD}. 
For $0.9 \lesssim J_1 \lesssim 0.95$, the increase of $H$ in the DT-SkX phase leads to the phase transition to the S-SkX. 
Both $M^z$ and $\chi^{\rm sc}$ show jumps in the transition, as shown in Fig.~\ref{fig: M-H}(b). 
In contrast to the DT-SkX, the skyrmion core aligns in a square-lattice way, as shown by the real-space spin configuration in the first row of Fig.~\ref{fig: SkX}(b). 
Accordingly, the distribution of the scalar chirality looks fourfold symmetric, as shown in the second row in Fig.~\ref{fig: SkX}(b). 
The square-lattice symmetry is also found in the spin structure factor in the third and fourth rows in Fig.~\ref{fig: SkX}(b); the largest intensity is found at $\bm{Q}_1$ and $\bm{Q}_2$, while the second largest one is found at $\bm{Q}_7$ and $\bm{Q}_8$. 
On the other hand, there are almost no intensities at $\bm{Q}_3$--$\bm{Q}_6$.
Thus, the S-SkX is stabilized by the interplay between the momentum-resolved interactions at $\bm{Q}_1$, $\bm{Q}_2$, $\bm{Q}_7$, and $\bm{Q}_8$. 
The relations as $\bm{Q}_1+\bm{Q}_2-\bm{Q}_7=\bm{0}$ and $\bm{Q}_1-\bm{Q}_2-\bm{Q}_8=\bm{0}$ induce the instability toward the S-SkX, which has been also found in similar tetragonal magnets~\cite{Hayami_PhysRevB.105.174437, hayami2022multiple, hayami2023widely}.

Compared to the DT-SkX, the S-SkX exhibits a smaller spin chirality, as shown in Fig.~\ref{fig: M-H}(b). 
This is intuitively understood from the difference in the skyrmion density, which is determined by the ordering wave vectors. 
As shown in the first row of Figs.~\ref{fig: SkX}(a) and \ref{fig: SkX}(b), the number of the skyrmion core in the 12$\times$12 spins is eight for the DT-SkX, while that is four for the S-SkX. 
In addition, one finds that the magnetization for the DT-SkX is smaller than that for the S-SkX, since the number of the skyrmion core with $S_i^z=-1$ is larger for the former SkX. 
As a result, the S-SkX is stabilized in the higher-field region compared to the DT-SkX. 

Let us discuss the relevant materials in experiments. 
The phase transition from the DT-SkX to the S-SkX against $H$ has been found in a polar tetragonal compound EuNiGe$_3$~\cite{ryan2016complex, Fabreges_PhysRevB.93.214414, singh2023transition, matsumura2023distorted}, where the recent small-angle neutron scattering measurements have revealed the emergence of the structural phase transition in terms of the SkX phases~\cite{singh2023transition}. 
In this material, a similar model with competing exchange interactions in momentum space exhibits such a phase transition. 
From the fact that the positions of the ordering wave vectors are different between the previous and present models, one finds that magnetic frustration in momentum space commonly plays an important role in inducing multiple SkX phases.

When $J_1$ is larger than $J_2$, the DT-SkX is no longer stabilized. 
Instead of the DT-SkX, the DS-SkX appears in the phase diagram in Fig.~\ref{fig: PD}. 
The spin and scalar chirality configurations of the DS-SkX are shown in the first and second rows of Fig.~\ref{fig: SkX}(c), respectively. 
Although the skyrmion core aligns in the square-lattice way, which is similar to the S-SkX, it is elongated along the [110] direction. 
Indeed, the intensities of the spin structure factor at $\bm{Q}_7$ and $\bm{Q}_8$ are different from each other, as shown in the third and fourth rows of Fig.~\ref{fig: SkX}(c). 
Since the DS-SkX turns into the S-SkX as $H$ increases, this phase transition is also regarded as the phase transition in terms of the different types of SkXs. 
In contrast to the phase transition between the DT-SkX and S-SkX, the magnetization and scalar chirality continuously behave in the phase transition between the DS-SkX and S-SkX, as shown in Fig.~\ref{fig: M-H}(c). 
A similar phase sequence has also been found in the centrosymmetric tetragonal compound EuAl$_4$, where the importance of the competing interactions in momentum space was pointed out~\cite{takagi2022square, hayami2023orthorhombic}. 
The present result indicates that the momentum-resolved DM interaction also becomes a source of inducing multiple SkX phases in noncentrosymmetric magnets. 

For $J_1>J_2$, the 1$Q$ CS and 2$Q'$ I states remain stable in the low- and high-field regions, respectively. 
Meanwhile, the 3$Q'$ state is replaced by the 2$Q'$ II state, whose spin configuration is characterized by the $\bm{Q}_1$, $\bm{Q}_2$, $\bm{Q}_7$, and $\bm{Q}_8$ components in the spin moments. 
Similarly to the 3$Q'$ state, there is no almost scalar chirality in this state, as shown in Fig.~\ref{fig: M-H}(c).

\section{Summary}
\label{sec: Summary}

To summarize, we have investigated the structural phase transitions in terms of the SkX spin textures in polar tetragonal magnets. 
By focusing on the role of magnetic frustration in momentum space, we obtain three types of SkX phases at low temperatures by performing the simulated annealing. 
Especially, we find two characteristic phase transitions when the magnetic field changes: One is the transition between the DT-SkX and S-SkX and the other is the transition between the DS-SkX and S-SkX. 
Our results indicate the importance of competing interactions in momentum space, whose mechanism is related to that causing similar structural SkX phase transitions in both noncentrosymmetric magnet EuNiGe$_3$ and centrosymmetric magnet EuAl$_4$. 
Based on this mechanism, further intriguing transitions with respect to the topological spin textures are expected.

\begin{acknowledgments}
The author thanks J. S. White, D. Singh, and N. Kanazawa for fruitful discussions. 
This research was supported by JSPS KAKENHI Grants Numbers JP21H01037, JP22H04468, JP22H00101, JP22H01183, JP23H04869, JP23K03288, and by JST PRESTO (JPMJPR20L8) and JST CREST (JPMJCR23O4).  
Parts of the numerical calculations were performed in the supercomputing systems in ISSP, the University of Tokyo.
\end{acknowledgments}

\appendix

\bibliographystyle{apsrev}
\bibliography{../ref.bib}
\end{document}